\documentclass[aps,prd,secnumarabic,nobibnotes,twocolumn,superscriptaddress]{revtex4-1}
\usepackage{amsfonts}
\usepackage{mathrsfs}
\usepackage{amsmath}% needed for subequations
\usepackage{color}
\usepackage{natbib}
\usepackage{graphicx}
\usepackage{bm}% bold maths
\usepackage{amssymb}
\usepackage{xspace}
\usepackage{epstopdf}
\usepackage{dcolumn}
\usepackage{multirow}
\usepackage[colorlinks=true, letterpaper=true, pdfstartview=FitV, linkcolor=blue, citecolor=blue, urlcolor=blue]{hyperref}
\usepackage{wrapfig}

\makeatletter

\newcommand{\Rmnum}[1]{\expandafter\@slowromancap\romannumeral #1@}
\makeatother

\begin{document}
\title{Recipe for single-pair-Weyl-points phonons carrying the same chiral charges}

\author{Guangqian Ding}\thanks{These authors contributed equally to this manuscript.}
\address{School of Science, Chongqing University of Posts and Telecommunications, Chongqing 400065, China}
\address{School of Physical Science and Technology, Southwest University, Chongqing 400715, China}

\author{Chengwu Xie}\thanks{These authors contributed equally to this manuscript.}
\address{School of Physical Science and Technology, Southwest University, Chongqing 400715, China}

\author{Jingbo Bai}
\address{School of Physical Science and Technology, Southwest University, Chongqing 400715, China}

\author{Zhenxiang Cheng}\email{cheng@uow.edu.au}\thanks{Corresponding author.}
\address{Institute for Superconducting and Electronic Materials (ISEM), University of Wollongong, Wollongong 2500, Australia}

\author{Xiaotian Wang}\email{xiaotianwang@swu.edu.cn}\thanks{Corresponding author.}
\address{School of Physical Science and Technology, Southwest University, Chongqing 400715, China}

\author{Weikang Wu}\email{weikang$_$wu@sdu.edu.cn}\thanks{Corresponding author.}
\address{Key Laboratory for Liquid-Solid Structural Evolution and Processing of Materials (Ministry of Education), Shandong University, Jinan, Shandong 250061, China}

\begin{abstract}
Recently, Wang \textit{et al}. [Phys. Rev. B, \textbf{106}, 195129 (2022)] challenged a widely held belief in the field of Weyl physics, demonstrating that single-pair-Weyl-points (SP-WPs) can exist in nonmagnetic spinless systems, contrary to previous assumptions that they could only exist in magnetic systems. Wang \textit{et al}. observed that the SP-WPs with opposite and even chiral charges ($i.e.$, $|\mathcal{C}|$ = 2 or 4) could also exist in nonmagnetic spinless systems. In this Letter, we present a novel finding in which SP-WPs have a partner, namely a charged nodal surface, in nonmagnetic spinless systems. In contrast to previous observations, we show that the SP-WPs can have uneven chiral charges ($i.e.$, $|\mathcal{C}|$ = 1). We identify 6 (out of 230) space groups (SGs) that contain such SP-WPs by searching the encyclopedia of emergent particles in three-dimensional crystals. Our finds were confirmed through the phonon spectra of two specific materials Zr$_3$O (with SG 182) and NaPH$_2$NO$_3$ (with SG 173). This discovery broadens the range of materials that can host SP-WPs and applies to other nonmagnetic spinless crystals.
\end{abstract}
\maketitle

%%%%%%% Main text %%%%%%%%%%%%%%%%%%%%%
\textcolor{blue}{\textit{Introduction.}} An outstanding challenge in Weyl physics is to realize an ideal Weyl state with a minimal number of Weyl points (WPs), typically a single pair~\cite{add1,add2}. The presence of minimum WPs is due to the simplicity it provides in studying their physical properties and imaging surface state modes in spectroscopy experiments~\cite{add3}. Theoretical models and experiments are simplified when the number of WPs is kept to a minimum. In the past, it was commonly believed that single-pair-Weyl-points (SP-WPs) could only be realized in magnetic systems with broken time-reversal symmetry. The statement appeared in the pioneering work by Wan \textit{et al}.~\cite{add4} and was cited in subsequent works~\cite{add3,add5,add6,add7,add8,add9}. Consequently, the search for SP-WPs was focused on magnetic materials in the past decade, with limited success: To date, there are only a few candidates, such as MnBi$_2$Te$_4$~\cite{add10,add11}, EuCd$_2$As$_2$~\cite{add12,add13,add14}, and their band structures are far from ideal.

In 2022, Wang \textit{et al}.~\cite{add15} discovered a loophole in Wan \textit{et al}'s original argument~\cite{add4}, demonstrating that a single pair of Weyl points can exist in nonmagnetic systems. Specifically, Wang \textit{et al}.~\cite{add15} showed that the SP-WPs could exist at two time-reversal-invariant momenta points in nonmagnetic spinless systems. Moreover, they~\cite{add15} proved that the two WPs of the SP-WPs have opposite and even charges ($i.e.$, $|\mathcal{C}|$ = 2 or 4). Following this breakthrough, Li \textit{et al}.~\cite{add16} experimentally observed SP-WPs with $|\mathcal{C}|$ = 2 (specifically, one WP with $|\mathcal{C}|$ = 2 and the other one with $|\mathcal{C}|$ = $-$2) in artificial periodic systems. Moreover, Wang \textit{et al}.~\cite{add17} proposed an ideal artificial periodic system to realize the SP-WPs with $|\mathcal{C}|$ = 4 (specifically, one WP with $|\mathcal{C}|$ = 4 and the other one WP with $|\mathcal{C}|$ = $-$4), noting that the charges in WPs have an upper limit and that their maximum charge should be four ($|\mathcal{C}|$ = 4)~\cite{add18,add19,add20,add21,add22}.

\textit{One may wonder if each WPs of the SP-WPs at the high-symmetry points (HSPs) can have the same and uneven charge.} In this work, we demonstrate that it is possible to achieve two WPs with a same and uneven charge in the nonmagnetic spinless systems if the SP-WPs have a partner, $i.e.$, a charged nodal surface state. To achieve such SP-WPs, we searched the encyclopedia of emergent particles in 3D crystals~\cite{add18} and developed a strategy. We find that SP-WPs at the HSPs (with a same and uneven charge) can be hosted in 6 (out of 230) space groups (SGs), and we confirm that the $|\mathcal{C}|$ of each WP is equal to 1. We then demonstrate our approach by presenting the phonon dispersions for two realistic material candidates Zr$_3$O (with SG 182) and NaPH$_2$NO$_3$ (with SG 173). It is worth noting that phonons~\cite{add23,add24,add25} play a vital role in many solid-state system features, including thermal and electrical conductivity, neutron scattering, and related phenomena such as superconductivity. Consequently, it is hoped that topological phonons~\cite{add26,add27,add28,add29,add30,add31,add32,add33,add34,add35,add36,add37,add38,add39,add40,add41,add42,add43,add44,add45,add46} will lead to rich and unconventional physics, and the search for materials containing topological phonons will become a key priority in the field.

In both materials, SP-WPs (with $|\mathcal{C}|$ = $+1$ for Zr$_3$O and with $|\mathcal{C}|$ = $-1$ for NaPH$_2$NO$_3$), along with a partner charged nodal surface state (with $|\mathcal{C}|$ = $-2$ for Zr$_3$O and with $|\mathcal{C}|$ = $+2$ for NaPH$_2$NO$_3$), appear in their phonon spectrums. Note that all the phonon bands are relevant for experimental detection, as the phonons are not subject to the Pauli exclusion principle and Fermi surface constraints. In addition to achieving two WPs with a same charge, these systems have another advantage~\cite{add47,add48}: A charged nodal surface state appears at the $k_z =\pm\pi$ planes in the Brillouin zone (BZ), allowing investigation of the relationship between the charged WPs and the charged nodal surface in a single system. Our proposal applies to spinless systems ranging from phononic to other bosonic systems, and even to electronic systems without spin-orbit coupling.

\begin{table*}
\centering
\renewcommand\arraystretch{1.5}
\caption{\label{Tab1} The candidate SGs that can host C-1 SP-WPs (with the same charge) in spinless systems. Irreps denotes the irreducible (co-) representation of the little group associated with the two C-1 WP at the K and K' HSPs. The symbols of the Irreps are adapted from Refs.~\cite{add18,add55}.}
\begin{ruledtabular}
\begin{tabular}{cccccc}
  SG  &   Generators                                                              &  HSPs  &  Irreps    &  Species & Materials \\
  169 &  $\{C_{3}^{+}|00\frac{1}{3}\}$,$\{C_{6}^{+}|00\frac{1}{6}\}\mathcal{T}$  &  K; K'  & $R_2, R_3$  &  C-1 WP &  \\
  170 &  $\{C_{3}^{+}|00\frac{1}{3}\}$,$\{C_{6}^{+}|00\frac{1}{6}\}\mathcal{T}$  &  K; K'  & $R_2, R_3$  &  C-1 WP &  Tl$_3$PO$_4$, Tl$_3$AsO$_4$, \\
      &                                                                          &         &             &         & Y$_3$CuGeS$_7$, NaPH$_2$NO$_3$ \\
  173 &  $\{C_{3}^{+}|000\}$,$\{C_{6}^{+}|00\frac{1}{2}\}\mathcal{T}$            &  K; K'  & $R_2, R_3$  &  C-1 WP &   \\
  178 &  $\{C_{3}^{+}|00\frac{1}{3}\}$,$\{C_{21}^{''}|00\frac{1}{2}\}$,$\{C_{6}^{+}|00\frac{1}{6}\}\mathcal{T}$  &  K; K'  & $R_3$  &  C-1 WP  & Hf$_5$Ir$_3$ \\
  179 &  $\{C_{3}^{+}|00\frac{2}{3}\}$,$\{C_{21}^{''}|00\frac{1}{2}\}$,$\{C_{6}^{+}|00\frac{5}{6}\}\mathcal{T}$  &  K; K'  & $R_3$  &  C-1 WP  &  \\
  182 &  $\{C_{3}^{+}|000\}$,$\{C_{21}^{''}|00\frac{1}{2}\}$,$\{C_{6}^{+}|00\frac{1}{2}\}\mathcal{T}$            &  K; K'  & $R_3$  &  C-1 WP  & ReO$_3$, Zr$_3$O \\
\end{tabular}\end{ruledtabular}
\end{table*}

\textcolor{blue}{\textit{General analysis.}} Based on the Nielsen-Ninomiya no-go theorem~\cite{add49,add50}, Weyl systems must satisfy compensation effects. This means that there must be at least one pair of Weyl points (WPs) with opposite charges in the BZ, and the total chiral charge of all the WPs should be zero. Thus, two WPs with the same charge ($i.e.$, all the WPs have a net chiral charge) cannot exist alone in the BZ. In order to eliminate the net chiral charge in BZ, the two WPs (with a same charge) need a partner in the BZ.

\begin{figure}[t]
\includegraphics[width=8cm]{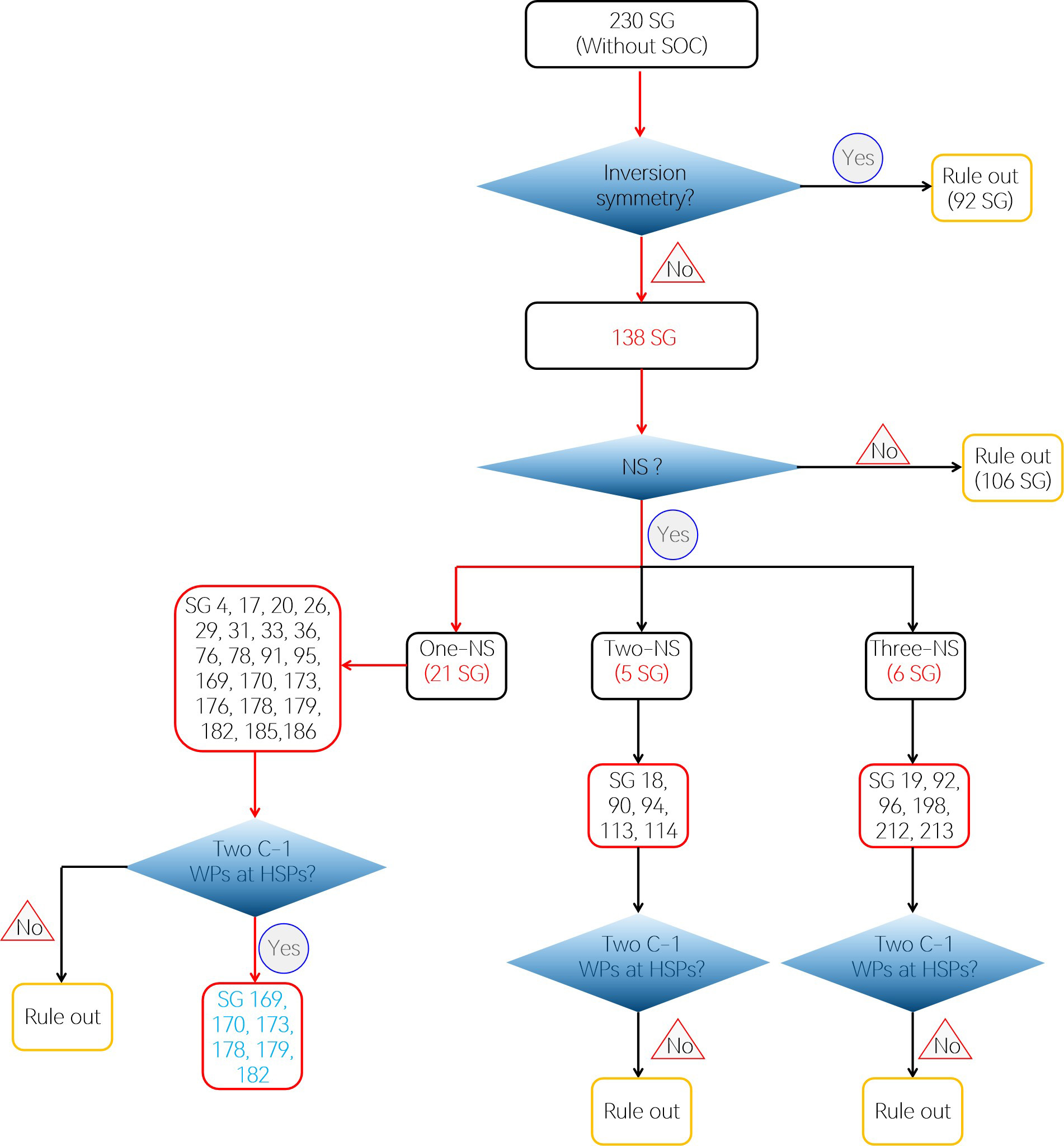}
\caption{Flowchart of screening for SP-WPs with the same $\mathcal{C}$ in spinless systems from the encyclopedia of emergent particles in Ref.~\cite{add18}
\label{fig1}}
\end{figure}

In this work, we focus on spinless systems and consider charged nodal surfaces as the partner of the two WPs. First, we list $three$ $symmetry$ $conditions$ that can help to realize SP-WPs with charged nodal surfaces: (i) The space group (SG) must allow for the coexistence of WPs and nodal surfaces in a same system; (ii) To achieve a nonzero topological charge for nodal surfaces, the system should be noncentrosymmetric~\cite{add51,add52,add53,add54}, as inversion ($\mathcal{P}$) together with time-reversal ($\mathcal{T}$) symmetry leads to the zero Chern number; (iii) To ensure SP-WPs at the high-symmetry points (HSPs) sharing the same $\mathcal{C}$, the pair of WPs should be related to each other by a proper rotation or $\mathcal{T}$. In other words, for the little group $S$ of a HSP as a subgroup of the SG $G$, the representative of coset decomposition of $G$ over $S$ should consist only of proper rotations and/or $\mathcal{T}$, not merely the identity. Taking into account the (i)-(iii) conditions, we screened the encyclopedia of emergent particles in spinless 3D crystals~\cite{add18}, following the strategy shown in Fig.~\ref{fig1}.

 We found that 138 candidate SGs lacked inversion symmetry. Among these, 21 SGs, 5 SGs, and 6 SGs were identified as candidate SGs to host one-, two-, and three-nodal surface states~\cite{add55,add56,add57,add58,add59,add60}, respectively. Based on the work of Wang \textit{et al}.~\cite{add15}, if the SP-WPs (in spinless systems) have no partner in the 3D BZ, meaning that all coexisting band degeneracies in the 3D BZ are WPs, the charges for each Weyl point in the SP-WPs must be even ($|\mathcal{C}|$ = 2 or 4) and opposite. Additionally, according to the work of Wang \textit{et al}. and Liu \textit{et al}.~\cite{add61,add62}, the C-3 WP (in spinless systems) can only appear on the high-symmetry paths, not at the HSPs. Therefore, in this Letter, we focus on the case of C-1 WPs at the HSPs, which host the same charge $\mathcal{C}$ = 1 or -1.
\begin{figure*}[t]
\includegraphics[width=13cm]{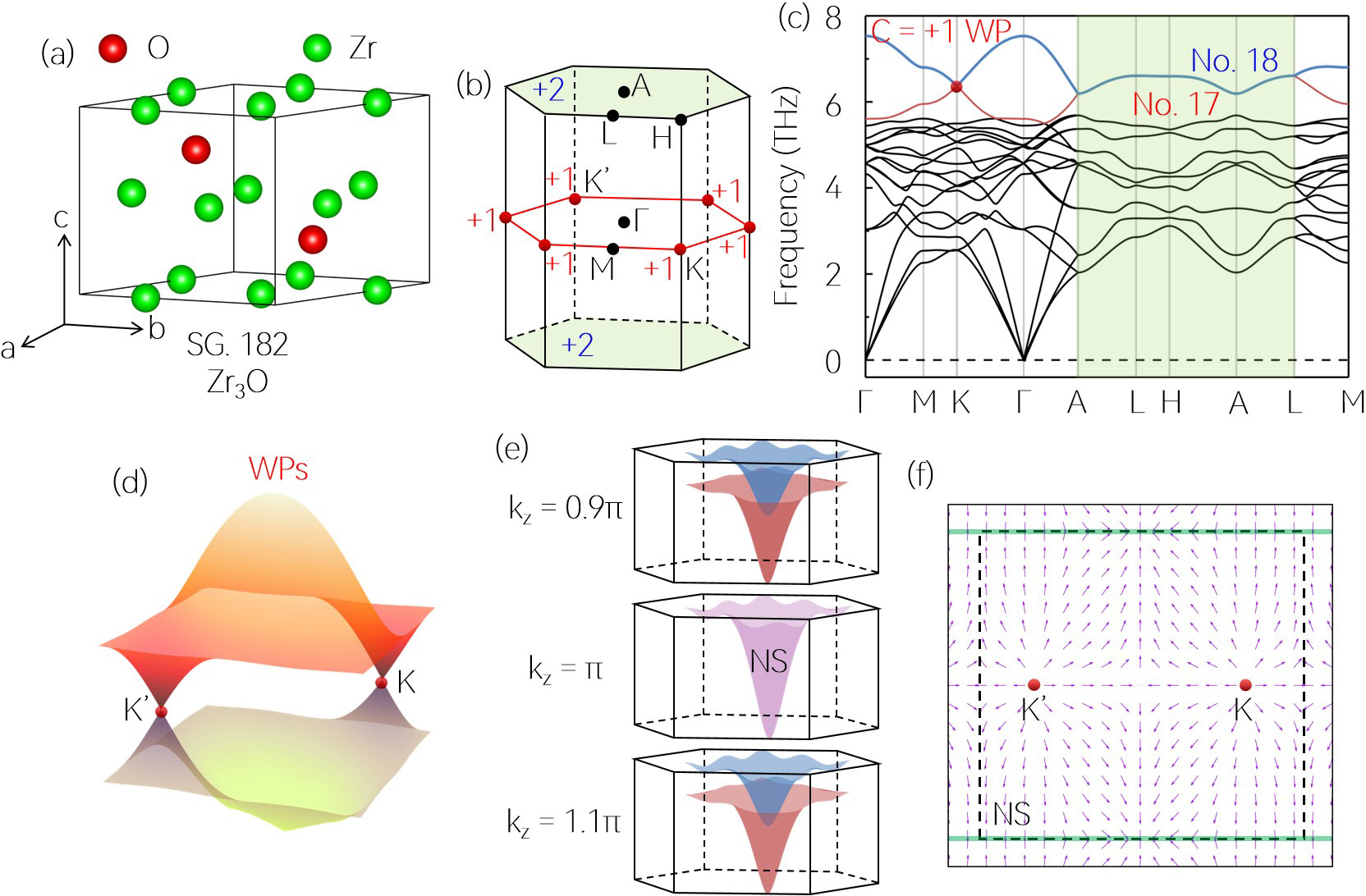}
\caption{(a) Crystal structure for Zr$_3$O with SG $P6_322$ (No. 182). (b) 3D BZ and the positions of one-nodal surface (NS) and two WPs with the same charge. (c) phonon dispersion for Zr$_3$O. The phonon branches No. 17 and No. 18 form a doubly degenerate point (with a red dot) at the K and a doubly degenerate line along the A-L-H-A paths (with green background). (d) 3D plot of the phonon bands around the SP-WPs at the K and K'. (e) 3D plots of the phonon bands at the $k_z$ = 0.9 $\pi$, $k_z$ = $\pi$, and $k_z$ = 1.1 $\pi$ planes. A nodal surface (NS) can be found at the $k_z$ = $\pi$ plane. (f) Distribution of Berry curvature in the $k_x$ = 0 plane. The dashed line indicates the first BZ, and the green line shows the NS.
\label{fig2}}
\end{figure*}

\begin{figure*}[t]
\includegraphics[width=10cm]{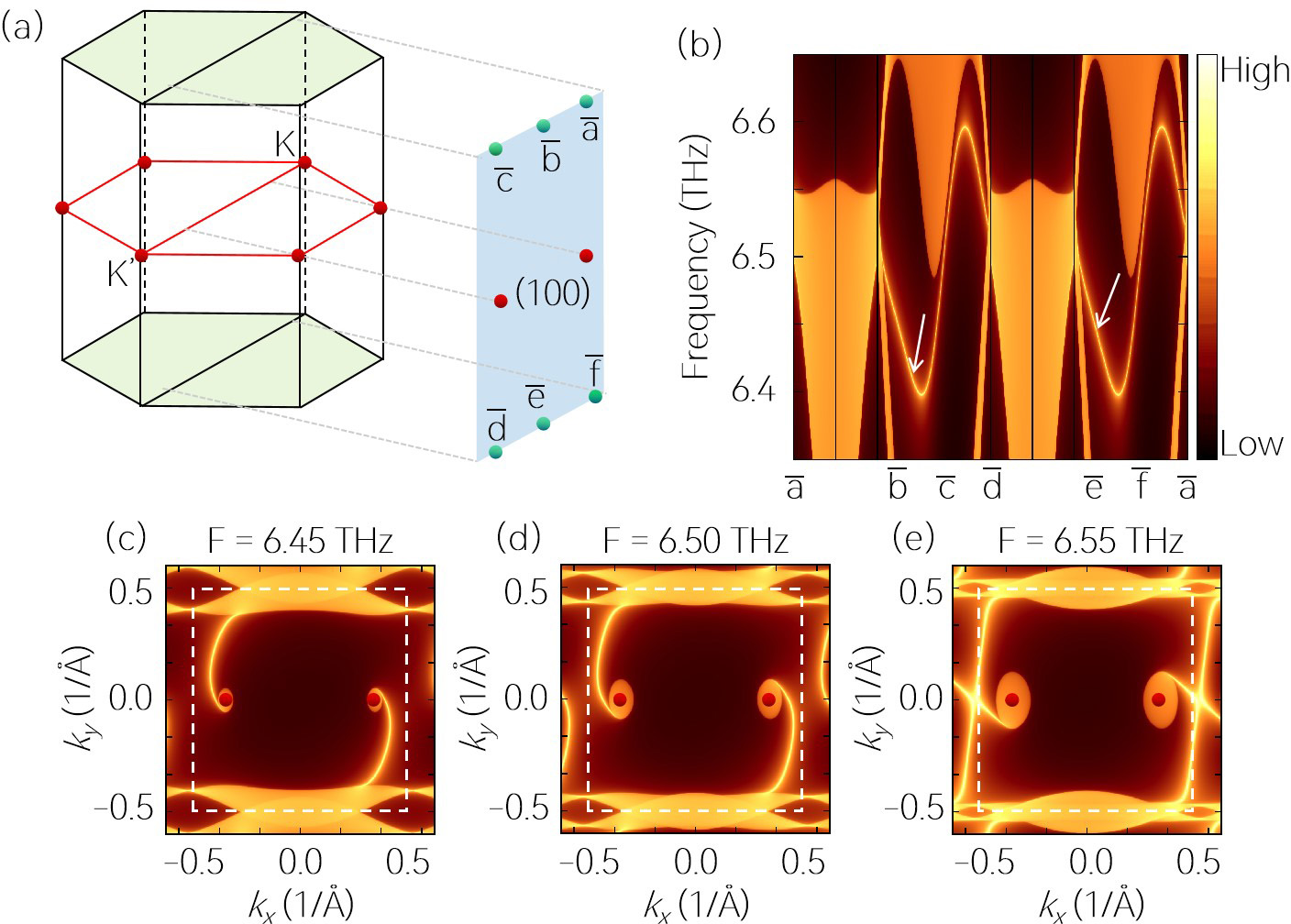}
\caption{(a) 3D bulk and 2D surface BZs. (b) phonon LDOS projected on the (100) surface for Zr$_3$O along the $\bar{a}-\bar{b}-\bar{c}-\bar{d}-\bar{e}-\bar{f}-\bar{a}$ surface paths. (c)-(e) isofrequency surface contours of the (100) surface at the frequencies of 6.45 THz, 6.50 THz, and 6.55 THz, respectively.
\label{fig3}}
\end{figure*}

From Fig.~\ref{fig1}, one can see that the two C-1 WPs do not exist at the HSPs, along with the two- and three-nodal surfaces in the spinless systems. Only six candidate SGs, with Nos. 169, 170, 173, 178, 179, and 182 can cohost charged one-nodal surface and two C-1 WPs (with the same charge) at the HSPs in spinless systems. The results are presented in Table~\ref{Tab1}, which lists the candidate SGs, the possible locations of the two WPs, and the corresponding symmetry representation. Note that the SP-WPs with the same charge are located at the HSPs K and K', and the topological charge $|\mathcal{C}|$ for the two WPs is noneven and equal to 1. Table I provides a recipe for the SP-WPs carrying the same chiral charges in spinless systems. In this main text, we shall take phonons in two real materials as examples. More details about the computational methods and the other material candidates can be found in the supplementary material (SM)~\cite{add63}.

\begin{figure*}[t]
\includegraphics[width=13cm]{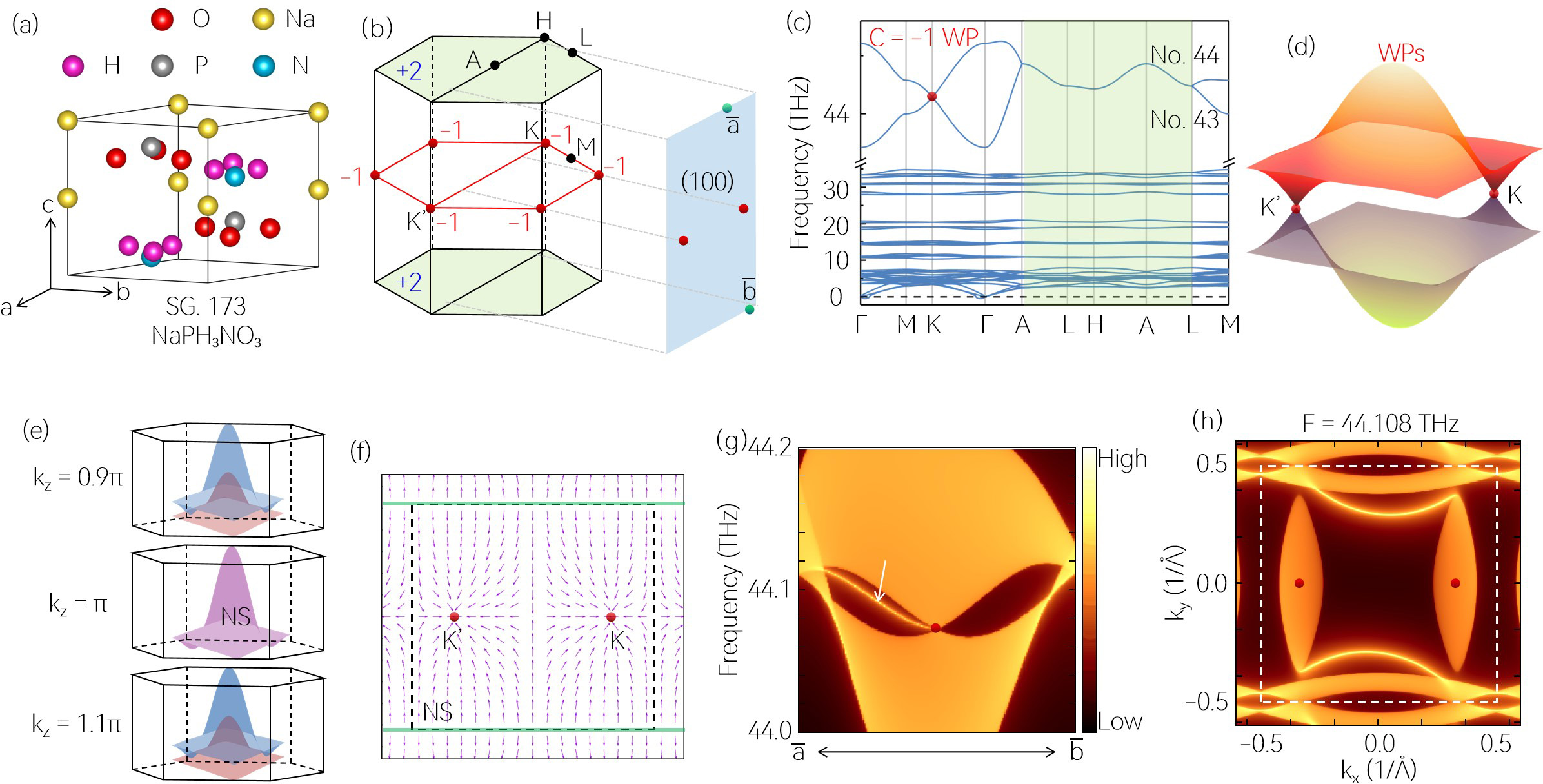}
\caption{(a) Crystal structure for NaPH$_2$NO$_3$ with SG P6$_3$ (No. 173). (b) 3D bulk and 2D surface BZs. (c) phonon dispersion for Zr$_3$O. The phonon branches No. 43 and No. 44 form a doubly degenerate point (with a red dot) at the K and a doubly degenerate line along the A-L-H-A paths (with green background). (d) 3D plot of the phonon bands around the SP-WPs at the K and K'. (e) 3D plots of the phonon bands at the $k_z$ = 0.9 $\pi$, $k_z$ = $\pi$, and $k_z$ = 1.1 $\pi$ planes. A nodal surface (NS) can be found at the $k_z = \pi$ plane. (f) Distribution of Berry curvature in the $k_x = 0$ plane. The dashed line indicates the first BZ. (g) Phonon LDOS projected on the (100) surface for NaPH$_2$NO$_3$ along the $\bar{a}$-$\bar{b}$ surface path. (h) isofrequency surface contour of the (100) surface at the frequency of 44.108 THz.
\label{fig4}}
\end{figure*}

\textcolor{blue}{\textit{Example 1: Zr$_3$O with SG 182.}} We show the first material example, Zr$_3$O with SG $P6_322$ (No. 182). The crystal structure for Zr$_3$O is shown in Fig.~\ref{fig2}(a), in which the Zr and O occupy the 6g, and 2c Wyckoff positions, respectively. The determined lattice constants from the first-principle calculations are a = b = 5.66 \AA, c = 5.23 \AA, which is in good agreement with the experimental ones~\cite{add64} (a = b = 5.61 \AA, c = 5.18 \AA). The phonon dispersion for Zr$_3$O along the $\Gamma$-M-K-$\Gamma$-A-L-H-A-L-M high-symmetry paths is shown in Fig.~\ref{fig2}(c). A supercell of 2$\times$2$\times$2 is adopted for the calculation of force constants. According to Fig.~\ref{fig2}(c), there is no imaginary frequency in the phonon spectrum, showing that Zr$_3$O is dynamically stable. We then focus on the frequency region of phonon branches No. 17 and No. 18 in Fig.~\ref{fig2}(c); one finds these two phonon branches are doubly degenerate at the HSP K and along the A-L-H-A paths. Actually, the degenerate point at the K is a WP, and one more WP can also be found at the HSP K' (see Fig.~\ref{fig2}(d)). The degenerate lines along the A-L-H-A (with green background) belong to a one-nodal surface at the $k_z = \pm\pi$ planes.

To better view the one-nodal surface, we show the 3D plot of the phonon bands for the $k_z$ = 0.9 $\pi$, $k_z$ = $\pi$, and $k_z$ = 1.1 $\pi$ planes in Fig.~\ref{fig2}(e). Obviously, the two phonon bands overlap entirely with each other and form a doubly degenerate band at the $k_z = \pi$ plane, reflecting the one-nodal surface behavior. The nodal surface~\cite{add59} is dictated by the combination of the time-reversal ($\mathcal{T}$) and twofold screw axis ($S_{2z}$) symmetries. At the $k_z = \pi$ plane, one notes $(S_{2z}\mathcal{T})^2 = -1$ which leads to the Kramers degeneracy over the whole plane. Away from this plane, the degeneracy is generally lifted due to the violation of $S_{2z}\mathcal{T}$, so that a nodal surface is formed at the $k_z = \pi$ plane. Note that the two WPs with the same charge and the one-nodal surface are topologically nontrivial. The two WPs (at the HSPs K and K') and the one-nodal surface (at the $k_z = \pm\pi$ planes) carry a nonzero topological charge, with $\mathcal{C} = +1$ for the WPs and $\mathcal{C} = -2$ for the one-nodal surface. This can be visualized in the plot of the Berry curvature field in Fig.~\ref{fig2}(f). One can see that the field is emitted from the two WPs with $\mathcal{C} = +1$ and absorbed by the one-nodal surface with $\mathcal{C} = -2$. Moreover, Zr$_3$O hosts a neutral chiral charge; the two WPs and their partner, a one-nodal surface, are constrained by the no-go theorem~\cite{add49,add50}. Note that the two WPs in the phonon dispersion of Zr$_3$O have the same and uneven charge, and the results of the first-principle calculations are in consistent with the results of the general analysis collected in Table~\ref{Tab1}.

Figure~\ref{fig3}(b) shows the phonon local density of states (LDOS) projected on the (100) surface for Zr$_3$O. Surface states in the phonon spectrum are clearly visible in the 6.45-6.6 THz frequency ranges. Figs.~\ref{fig3}(c)-(e) show isofrequency surface contours of the (100) surface at the frequencies of 6.45 THz, 6.50 THz, and 6.55 THz, respectively. These figures reveal two surface arcs connecting the projected nodal surface, each connecting to one of the projected SP-WPs (marked by red dots). These clean arcs are not masked by the bulk states and should be easily detected experimentally~\cite{add65}.

\textcolor{blue}{\textit{Example 2: NaPH$_2$NO$_3$ with SG 173.}} The second material example is NaPH$_2$NO$_3$ with SG P6$_3$ (No. 173). Figure~\ref{fig4}(a) depicts the relaxed crystal structure of NaPH$_2$NO$_3$; the Na, P, H, N, and O atoms occupy the 2a, 2b, 6c, 2b, and 6c Wyckoff positions, respectively. The lattice constants for NaPH$_2$NO$_3$ obtained in theory are a = b = 5.80 \AA, c = 6.16 \AA, agreeing well with the experimental ones~\cite{add66} (a = b = 5.77 \AA, c = 6.03 \AA). The phonon dispersion for NaPH$_2$NO$_3$ along the $\Gamma$-M-K-$\Gamma$-A-L-H-A-L-M paths is shown in Fig.~\ref{fig4}(c). From Fig.~\ref{fig4}(c), one finds that the one-nodal surface appears at the $k_z = \pi$ plane (see Fig.~\ref{fig4}(e)), and two WPs appear at the HSPs K and K' (see Fig.~\ref{fig4}(d)). The SP-WPs at the K and K' and the one-nodal surface carry a $\mathcal{C} = -1$ and $+2$, respectively. The corresponding Berry curvature field distribution is shown in Fig.~\ref{fig4}(f). Obviously, the Berry curvature near the one-nodal surface and the SP-WPs indicates convergent and divergent field morphology, respectively, corresponding to the chirality of one-nodal surface and the SP-WPs, respectively. Fig.~\ref{fig4}(g) shows the phonon LDOS projected on the (100) surface for NaPH$_2$NO$_3$. Furthermore, Fig.~\ref{fig4}(h) shows the isofrequency surface contour of the (100) surface at the frequency of 44.108 THz, respectively. As with Zr$_3$O, two surface arcs connect the projected nodal surface and the projected SP-WPs.

\textcolor{blue}{\textit{Summary.}} The SP-WPs previously proposed by Wang \textit{et al}.~\cite{add15} in spinless systems do not have a partner in the 3D BZ. In their work, the two WPs of the SP-WPs~\cite{add16} have opposite and even charges. In this Letter, we extend the previously proposed SP-WPs to a new case in the spinless systems, where they have a partner, i.e., a charged one-nodal surface, in the 3D BZ. In this case, the two WPs of the SP-WPs have the same and uneven charge, $i.e.$, $|\mathcal{C}|$ = 1. We proposed a recipe for the new case of SP-WPs based on the encyclopedia of emergent particles and reported that they could exist in 6 (out of 230) SGs, $i.e.$, 169, 170, 173, 178, 179, and 182. Based on the recipe, we also contribute to the materials' realization. The SP-WPs with $|\mathcal{C}|$ = 1 can be observed in the phonon spectra of two concrete materials: Zr$_3$O and NaPH$_2$NO$_3$. Our findings apply to many spinless systems, including phononic, other bosonic, and even electronic systems without spin-orbit coupling.


\begin{thebibliography}{alpha}
\bibitem{add1} B. Yan and C. Felser, Annu. Rev. Condens. Matter Phys. \textbf{8}, 337-354 (2017).
\bibitem{add2} N. P. Armitage, E. J. Mele, and A. Vishwanath, Rev. Mod. Phys. \textbf{90}, 015001 (2018).
\bibitem{add3} I. Belopolski, K. Manna, D. S. Sanchez, G. Chang, B. Ernst, J. Yin, S. S. Zhang, T. Cochran, N. Shumiya, and H. Zheng \textit{et al}., Science \textbf{365}, 1278 (2019).
\bibitem{add4} X. Wan, A. M. Turner, A. Vishwanath, and S. Y. Savrasov, Phys. Rev. B \textbf{83}, 205101 (2011).
\bibitem{add5} L. M. Schoop, F. Pielnhofer, and B. V. Lotsch, Chem. Mater. \textbf{30}, 3155 (2018).
\bibitem{add6} D. F. Liu, A. J. Liang, E. K. Liu, Q. N. Xu, Y. W. Li, C. Chen, D. Pei, W. J. Shi, S. K. Mo, P. Dudin, T. Kim, C. Cacho, G. Li, Y. Sun, L. X. Yang, Z. K. Liu, S. S. P. Parkin, C. Felser, and Y. L. Chen, Science \textbf{365}, 1282 (2019).
\bibitem{add7} S. Nie, T. Hashimoto, and F. B. Prinz, Phys. Rev. Lett. \textbf{128}, 176401 (2022).
\bibitem{add8} L. Lu, L. Fu, J. D. Joannopoulos, and M. Solja$\rm{\breve{c}}$i$\rm{\acute{c}}$, Nat. Photonics \textbf{7}, 294 (2013).
\bibitem{add9} O. Vafek and A. Vishwanath, Annu. Rev. Condens. Matter Phys. \textbf{5}, 83 (2014).
\bibitem{add10} J. Li, Y. Li, S. Du, Z. Wang, B.-L. Gu, S.-C. Zhang, K. He, W. Duan, and Y. Xu, Sci. Adv. \textbf{5}, eaaw5685 (2019).
\bibitem{add11} D. Zhang, M. Shi, T. Zhu, D. Xing, H. Zhang, and J. Wang, Phys. Rev. Lett. \textbf{122}, 206401 (2019).
\bibitem{add12} L. L. Wang, N. H. Jo, B. Kuthanazhi, Y. Wu, R. J. McQueeney, A. Kaminski, and P. C. Canfield, Phys. Rev. B \textbf{99}, 245147 (2019).
\bibitem{add13} J. R. Soh, F. de Juan, M. G. Vergniory, N. B. M. Schr$\rm{\ddot{o}}$ter, M. C. Rahn, D. Y. Yan, J. Jiang, M. Bristow, P. Reiss, J. N. Blandy, Y. F. Guo, Y. G. Shi, T. K. Kim, A. McCollam, S. H. Simon, Y. Chen, A. I. Coldea, and A. T. Boothroyd, Phys. Rev. B \textbf{100}, 201102(R) (2019).
\bibitem{add14} J.-Z. Ma, S. M. Nie, C. J. Yi, J. Jandke, T. Shang, M. Y. Yao, M. Naamneh, L. Q. Yan, Y. Sun, A. Chikina, V. N. Strocov, M. Medarde, M. Song, Y.-M. Xiong, G. Xu, W. Wulfhekel, J. Mesot, M. Reticcioli, C. Franchini, and C. Mudry \textit{et al}., Sci. Adv. \textbf{5}, eaaw4718 (2019).
\bibitem{add15} X. Wang, F. Zhou, Z. Zhang, W. Wu, Z.-M. Yu, and S. A. Yang, Phys. Rev. B \textbf{106}, 195129 (2022).
\bibitem{add16} X.-P. Li, D. Zhou, Y. Wu, Z.-M. Yu, F. Li, and Y. Yao, Phys. Rev. B \textbf{106}, L220302 (2022).
\bibitem{add17} Z. Q. Wang, Q. B. Liu, X. F. Yang, and H. H. Fu, Phys. Rev. B \textbf{106}, L161302 (2022).
\bibitem{add18} Z.-M. Yu, Z. Zhang, G.-B. Liu, W. Wu, X.-P. Li, R.-W. Zhang, S. A. Yang, and Y. Yao, Sci. Bull. \textbf{67}, 375 (2022).
\bibitem{add19} T. T. Zhang, R. Takahashi, C. Fang, and S. Murakami, Phys. Rev. B \textbf{102}, 125148 (2020).
\bibitem{add20} Q.-B. Liu, Y. Qian, H.-H. Fu, and Z. Wang, Npj Comput. Mater. \textbf{6}, 95 (2020).
\bibitem{add21} Q.-B. Liu, Z. Wang, and H.-H. Fu, Phys. Rev. B \textbf{103}, L161303 (2021).
\bibitem{add22} C. Cui, X. P. Li, D. S. Ma, Z. M. Yu, and Y. Yao, Phys. Rev. B \textbf{104}, 075115 (2021).
\bibitem{add23} G. Xie, D. Ding, and G. Zhang, Adv. Phys.: X \textbf{3}, 1480417 (2018).
\bibitem{add24} N. Li, J. Ren, L. Wang G. Zhang, P. H$\rm{\ddot{a}}$nggi, and B. Li, Rev. Mod. Phys. \textbf{84}, 1045 (2012).
\bibitem{add25} J. Chen, J. He, D. Pan, X. Wang, N. Yang, J. Zhu, S. A. Yang, and G. Zhang, Sci. China Phys. Mech. Astron. \textbf{65}, 117002 (2022).
\bibitem{add26} X. Wang, T. Yang, Z. Cheng, C. Surucu, J. Wang, F. Zhou, Z. Y. Zhang, and G. Zhang, Appl. Phys. Rev. \textbf{9}, 041304 (2022).
\bibitem{add27} Y. Liu, X. Chen, and Y. Xu, Adv. Funct. Mater. \textbf{30}, 1904784 (2020).
\bibitem{add28} S. Singh, Q. S. Wu, C. Yue, A. H. Romero, and A. A. Soluyanov, Phys. Rev. Materials \textbf{2}, 114204 (2018).
\bibitem{add29} J. Li, J. Liu, S. A. Baronett, M. Liu, L. Wang, R. Li, Y. Chen, D. Li, Q. Zhu, and X. Chen, Nat. Commun. \textbf{12}, 1204 (2020).
\bibitem{add30} B. Peng, S. Murakami, B. Monserrat and T. T. Zhang, Npj Comput. Mater. \textbf{7}, 195 (2021).
\bibitem{add31} Z. Huang, Z. Chen, B. Zheng, and H. Xu, Npj Comput. Mater. \textbf{6}, 87 (2020).
\bibitem{add32} J. J. Zhu, W. K. Wu, J. Z. Zhao, H. Chen, L. F. Zhang, and S. S. A. Yang, Npj Quantum Mater. \textbf{7}, 52 (2022).
\bibitem{add33} X.-Q. Chen, J. Liu, and J. Li, Innovation \textbf{2}, 100134 (2021).
\bibitem{add34} Y. Liu, N. Zou, S. Zhao, X. Chen, Y. Xu, and W. Duan, Nano Lett. \textbf{22}, 2120 (2022).
\bibitem{add35} H. Miao, T. T. Zhang, L. Wang, D. Meyers, A. H. Said, Y. L. Wang, Y. G. Shi, H. M. Weng, Z. Fang, and M. P. M. Dean, Phys. Rev. Lett. \textbf{121}, 035302 (2018).
\bibitem{add36} B. W. Xia, R. Wang, Z. J. Chen, Y. J. Zhao, and H. Xu, Phys. Rev. Lett. \textbf{123}, 065501 (2019).
\bibitem{add37} T. Zhang, Z. Song, A. Alexandradinata, H. Weng, C. Fang, L. Lu, and Z. Fang, Phys. Rev. Lett. \textbf{120}, 016401 (2018).
\bibitem{add38} C. Xie, Y. Liu, Z. Zhang, F. Zhou, T. Yang, M. Kuang, X. Wang, and G. Zhang, Phys. Rev. B \textbf{104}, 045148 (2021).
\bibitem{add39} W. W. Yu, Y. Liu, X. M. Zhang, and G. D. Liu, Phys. Rev. B \textbf{106}, 195142 (2022).
\bibitem{add40} Y. Feng, C. Xie, H. Chen, Y. Liu, and X. Wang, Phys. Rev. B \textbf{106}, 134307 (2022).
\bibitem{add41} W.-W. Yu, Y. Liu, W. Meng, H. Liu, J. Gao, X. Zhang, and G. Liu, Phys. Rev. B \textbf{105}, 035429 (2022).
\bibitem{add42} G. Liu, Y. Jin, Z. Chen, and H. Xu, Phys. Rev. B \textbf{104}, 024304 (2021).
\bibitem{add43} Z. J. Chen, R. Wang, B. W. Xia, B. B. Zheng, Y. J. Jin, Y. J. Zhao, and H. Xu, Phys. Rev. Lett. \textbf{126}, 185301 (2021).
\bibitem{add44} B. Zheng, B. Xia, R. Wang, Z. Chen, J. Zhao, Y. Zhao, and H. Xu, Phys. Rev. B \textbf{101}, 100303(R) (2020).
\bibitem{add45} B. Peng, Y. Hu, S. Murakami, T. Zhang, and B. Monserrat, Sci. Adv. \textbf{6}, eabd1618 (2020).
\bibitem{add46} J.-Y. You, X.-L. Sheng, and G. Su, Phys. Rev. B \textbf{103}, 165143 (2021).
\bibitem{add47} J. Wang, H. Yuan, M. Kuang, T. Yang, Z.-M. Yu, Z. Zhang, and X. Wang, Phys. Rev. B \textbf{104}, L041107 (2021).
\bibitem{add48} P. C. Sreeparvathy, C. Mondal, C. K. Barman, and A. Alam, Phys. Rev. B \textbf{106}, 085102 (2022).
\bibitem{add49} H. B. Nielsen and M. Ninomiya, Nucl. Phys. B \textbf{185}, 20 (1981).
\bibitem{add50} H. B. Nielsen and M. Ninomiya, Nucl. Phys. B \textbf{193}, 173 (1981).
\bibitem{add51} Y. Yang, J.-p. Xia, H.-x. Sun, Y. Ge, D. Jia, S.-q. Yuan, S. A. Yang, Y. Chong, and B. Zhang, Nat. Commun. \textbf{10}, 5185 (2019).
\bibitem{add52} G. Q. Chang, B. J. Wieder, F. Schindler, D. S. Sanchez, I. Belopolski, S. M. Huang, B. Singh, D. Wu, T. R. Chang, T. Neupert, S.-Y. Xu, H. Lin, and M. Z. Hassan, Nat. Mater. \textbf{17}, 978 (2018).
\bibitem{add53} M. Kim, D. Lee, D. Lee, and J. Rho, Phys. Rev. B \textbf{99}, 235423 (2019).
\bibitem{add54} M. Xiao, L. Ye, C. Qiu, H. He, Z. Liu, and S. Fan, Sci. Adv. \textbf{6}, eaav2360 (2020).
\bibitem{add55} C. Bradley and A. Cracknell, The Mathematical Theory of Symmetry in Solids: Representation Theory for Point Groups and Space Groups (Oxford University Press, 2010).
\bibitem{add56} C. Xie, H. Yuan, Y. Liu, and X. Wang, Phys. Rev. B \textbf{105}, 054307 (2022).
\bibitem{add57} C. Xie, H. Yuan, Y. Liu, X. Wang, and G. Zhang, Phys. Rev. B \textbf{104}, 134303 (2021).
\bibitem{add58} J. Wang, H. Yuan, Z.-M. Yu, Z. Zhang, and X. Wang, Phys. Rev. Mater. \textbf{5}, 124203 (2021)
\bibitem{add59} W. Wu, Y. Liu, S. Li, C. Zhong, Z.-M. Yu, X.-L. Sheng, Y. X. Zhao, and S. A. Yang, Phys. Rev. B \textbf{97}, 115125 (2018).
\bibitem{add60} X. M. Zhang, Z.-M. Yu, Z. M. Zhu, W. K. Wu, S.-S. Wang, X.-L. Sheng, and S. A. Yang, Phys. Rev. B \textbf{97}, 235150 (2018).
\bibitem{add61} X. T. Wang, F. Zhou, Z. Y. Zhang, Z. M. Yu, and Y. G. Yao, Phys. Rev. B \textbf{106}, 214309 (2022).
\bibitem{add62} G. Liu, Z. Chen, P. Wu, and H. Xu, Phys. Rev. B \textbf{106}, 214308 (2022).
\bibitem{add63} See the supplementary material (SM) for the computational details, and the other material candidates with single-pair-Weyl-points phonons carrying the same chiral charges, which includes Refs.~\cite{add67,add68,add69,add70,add71,add72,add73,add74,add75,add76}
\bibitem{add64} A. B. Riabov, V. A. Yartys, B. C. Hauback, P. W. Guegan, G. Wiesinger, and I. R. Harris, J. Alloys Compd., \textbf{93}, 293-295 (1999).
\bibitem{add65} T. T. Zhang, H. Miao, Q. Wang, J. Q. Lin, Y. Cao, G. Fabbris, A. H. Said, X. Liu, H. C. Lei, Z. Fang, H. M. Weng, and M. P. M. Dean, Phys. Rev. Lett. \textbf{123}, 245302 (2019).
\bibitem{add66} E. Hobbs, D. E. C. Corbridge and B. Raistrick, Acta Crystallographica, \textbf{123}, 621-626 (1953).
\bibitem{add67} G. Kresse and J. Hafner, Phys. Rev. B \textbf{49}, 14251 (1994).
\bibitem{add68} G. Kresse and J. Furthm$\rm{\ddot{u}}$ller, Phys. Rev. B \textbf{54}, 11169 (1996).
\bibitem{add69} P. E. Bl$\rm{\ddot{o}}$chl, Phys. Rev. B \textbf{50}, 17953 (1994).
\bibitem{add70} J. P. Perdew, K. Burke, and M. Ernzerhof, Phys. Rev. Lett. \textbf{77}, 3865 (1996).
\bibitem{add71} X. Gonze and C. Lee, Phys. Rev. B \textbf{55}, 10355 (1997).
\bibitem{add72} A. Togo and I. Tanaka, First principles phonon calculations in materials science, Scr. Mater.  \textbf{108}, 1 (2015).
\bibitem{add73} Q.-S. Wu, S.-N. Zhang, H.-F. Song, M. Troyer, and A. A. Soluyanov, Comput. Phys. Commun. \textbf{224}, 405 (2018).
\bibitem{add74} M. L. Sancho, J. L. Sancho, and J. Rubio, J. Phys. F \textbf{14}, 1205 (1984).
\bibitem{add75} M. P. L. Sancho, J. M. L. Sancho, and J. Rubio, J. Phys. F \textbf{15}, 851 (1985).
\bibitem{add76} A. A. Soluyanov and D. Vanderbilt, Phys. Rev. B \textbf{83}, 235401 (2011).

\end{thebibliography}
\end{document}